\documentstyle[12pt,aaspp4,epsf]{article}
\begin{document}
\newcommand{\gtsima}{$\; \buildrel > \over \sim \;$}
\newcommand{\ltsima}{$\; \buildrel < \over \sim \;$}
\newcommand{\simgt}{\lower.5ex\hbox{\gtsima}}
\newcommand{\simlt}{\lower.5ex\hbox{\ltsima}}
\newcommand{\himpc}{{\hbox {$h^{-1}$}{\rm Mpc}} }
\newcommand{\avr}[1]{\left\langle #1 \right\rangle}
\newcommand{\cum}[1]{\left\langle #1 \right\rangle_{\rm c}}
\newcommand{\bfm}[1]{{\mbox{\boldmath $#1$}}}
\newcommand{\sbfm}[1]{{\mbox{\footnotesize\boldmath $#1$}}}
\newcommand{\ssbfm}[1]{{\mbox{\scriptsize\boldmath $#1$}}}
\newcommand{\rot}{\mbox{rot}}
\newcommand{\mGamma}{{\mit\Gamma}}
\newcommand{\mOmega}{{\mit\Omega}}
\newcommand{\threej}[6]{\left(\begin{array}{ccc}
      #1 & #2 & #3 \\
      #4 & #5 & #6
     \end{array}\right)}
  \def\x{{\bf x}}
  \def\r{{\bf r}}
  \def\k{{\bf k}}
  \def\s{{\bf s}}
  \def\hx{{\hat\x}}
  \def\hr{{\hat\r}}
  \def\hk{{\hat\k}}
  \def\qq#1{{2 #1\kern-0.2em+\kern-0.2em1}}
  \def\hkr{\hk\hr}
  \def\hkx{\hk\hx}
  \def\eikr{e^{i\k\r}}
  \def\kint{ \int\kern-0.25em {d^3 k\over(2\pi)^3}\,}
  \def\pkint{\kint P(k)\,\eikr\,}
  \def\etal{{\it et al}\ }

\title{Redshift Space Distortions of the Correlation Function
          in Wide Angle Galaxy Surveys}

\author{Alexander S. Szalay}
\affil{Department of Physics and Astronomy,\\
     The Johns Hopkins University, Baltimore, MD 21218}

\author{Takahiko Matsubara}
\affil{Department of Physics, The University of Tokyo, Tokyo 113, Japan}

\author{Stephen D. Landy}
\affil{Department of Physics, College of William and Mary,
     Williamsburg, VA 23187-8795}

\begin{abstract}
Using a novel two-dimensional coordinate system, we have derived a
particularly simple way to express the redshift distortions in galaxy
redshift surveys with arbitrary geometry in closed form. This method
provides an almost ideal way to measure the value of
$\beta=\Omega_0^{0.6}/b$ in wide area surveys, since all pairs in the
survey can be used for the analysis. In the limit of small angles,
this result straightforwardly reduces to the plane-parallel
approximation.  This expansion can also be used together with more
sophisticated methods such as for the calculation of Karhunen-Loeve
eigenvectors in redshift space for an arbitrary survey
geometry. Therefore, these results should provide for more precise
methods in which to measure the large scale power spectrum and the
value of $\beta$ simultaneously.
\end{abstract}

\keywords{cosmology: large-scale structure of the universe}

\section{Introduction}

  The fact that the correlation function in redshift space is
distorted from the real space correlations is well known. Davis and
Peebles (1983) have demonstrated quantitatively that the correlation
function $\xi(\pi,\sigma)$, expressed as a function of the
line-of-sight $(\pi)$ and perpendicular $(\sigma)$ separations, has a
significant anisotropy. By measuring the elongation of this function
along the $(\pi)$ axis, a value of the one-dimensional velocity
dispersion can be inferred. They found that the velocity distribution
was well described by an exponential distribution and that the value
of the dispersion was relatively independent of scale, at around 350
km/s. Peebles (1980) gives a nice summary why in linear theory the
typical velocities of the galaxies scale with $\Omega_0^{0.6}$. With
the advent of biased galaxy formation (Bardeen \etal 1986), the
velocities have been rescaled by a bias factor $b$, leading to
velocities dependent upon the combination $\beta=\Omega^{0.6}/b$.

In Kaiser's (1987) pioneering work, the effect of the infall due to
linear theory was identified and it was shown that in the limit that
the lines of sight to the two galaxies are approximately parallel to
each other, the power spectrum $P(k)$ is enhanced as a function of the
directional cosine $\mu$ between the wave vector and the line-of-sight as
\begin{equation}
     P^s(k,\mu)  = P(k) (1+\beta \mu^2)^2.
\end{equation}

     In related work, Lilje and Efstathiou(1989) calculated the
angular average of the redshift correlation function directly related
to this expression.  Hamilton(1992) expanded the redshift space
correlation function into components, multiplied with the angular
multipoles.  He showed that only the quadrupole and hexadecapole terms
arise in linear theory, in agreement with Kaiser's expression.  Other
promising approaches have involved an expansion into orthogonal
eigenfunctions (Heavens and Taylor 1994), or restricting the surveys
to small opening angles so that the plane-parallel approximation still
holds (Cole, Fisher and Weinberg 1994,1995). These papers attempted to
measure the parameter $\beta$ from the quadrupole to monopole ratio.

Zaroubi and Hoffmann (1996) outlined how to compute a linear expansion
of the redshift space correlation function for a general geometrical
configuration and provided numerical estimates of the redshift
distortions.  Additionally, Hamilton and Culhane (1996), hereafter
HC96, have introduced a novel integral transform, rotationally
invariant and commuting with the redshift distortion operator, which
makes the transformation very elegant and simple.

Here we present simple and intuitive expressions to calculate the
redshift distortions which exploit the fact that the inherent geometry
of the problem is two dimensional.  Indeed, by rotating all pairs in a
redshift survey to a common plane and analyzing the redshift
distortions in that plane, we obtain particularly simple analytic
results. These results smoothly approach the Kaiser/Hamilton
plane-parallel limit in which angle between the galaxy pair in a
survey is small. Additionally, these expressions allow for a much more
productive use of redshift surveys since all galaxy pairs can be
utilized in a given wide angle survey. It can also be used in adjunct
with more sophisticated methods of analysis such as construction of
eigenvectors for a KL analysis (Vogeley and Szalay 1996).

\section{The Correlation Function in Redshift Space}

\subsection{The Coordinate System}

The symmetries in the geometric configuration of redshift surveys have
been elegantly described in HC96, who recognized that the correlation
function in redshift space is tied to a triangle formed by the
observer and the two galaxies which lie upon two arbitrary lines of
sight. The correlation function should be a function of this triangle,
but otherwise invariant under rotations about the observer. Such
invariant triangles can be characterized by one size and two shape
parameters. We will use two angles to describe the shape of the
triangle, and express the scale dependence of the correlation function
as a function of $r = |\r|$, the distance between the two points.

In this coordinate system, which is illustrated in Fig. 1, the two
points are located at points $\r_1 = r_1 \hr_1$ and
$\r_2=r_2\hr_2$. The observer is at the origin, at rest in comoving
frame. The angle between the two normal vectors is $2\theta$, that is
$\hr_1\hr_2 =\cos 2\theta$.  The separation between the two points is
$\r = \r_1-\r_2$, with a normal vector $\hr$.  The angles between
$\hr_1,\hr_2$ and $\hr$ are $\gamma_1$ and $\gamma_2$
respectively. The symmetry axis for $\hr_1$ and $\hr_2$ is found by
halving the angle $2\theta$. The angle between this symmetry axis and
$\r$ is $\gamma$.  We assume that $r_1>r_2$, thus $\gamma_2>\gamma_1$.
The angles are related to each other as $\gamma_1+\gamma_2 = 2\gamma$,
and $\gamma_2-\gamma_1 = 2\theta$.  This introduction implies our
choice of the two angles, $\gamma$ and $\theta$, which together with
$|\r|$ completely describe the shape of the triangle.  These angles
are particularly convenient since $\theta$ is given by our geometry
and $\gamma$ smoothly approaches the angle familiar in the
plane-parallel limit as $\theta$ tends to 0.  There is another
geometric relation between the two angles: if $\theta$ is not 0, then
the smallest value $\gamma$ can take is $\theta$, and its largest
value is $\pi-\theta$. This is not due to the choice of coordinates
but rather to the geometry of the problem.

We will provide an expression for the distortions of the correlation
function as a series expansion in terms of the angle $\gamma$, where
the coefficients have a simple $r$ and $\theta$ dependence, together
with a dependence on the parameter $\beta$ that we wish to measure.
So far our approach has been quite similar to most earlier work except
for the introduction of a convenient set of angles which will simplify
the results considerably.

It has been a persistent problem to get a simple power spectrum or
correlation function beyond the Kaiser (1987) plane-parallel limit.
The difficulty lies in that the Fourier space transfer function, the
ratio of the distorted and undistorted power spectrum contains a
strong mode-coupling (Zaroubi and Hoffmann 1996). Thus, the transfer
function is not multiplicative but contains an integral over a
non-local kernel with a strong $k$-dependence. This dependence
strongly affects the multipole coefficients which would otherwise make
a precise determination of $\beta$ quite straightforward. One
alternative has been to use only galaxy pairs at small angular
separations (e.g. Cole \etal 1994, 1995).

It is interesting to consider what the reasons are for this
non-locality. With a finite $\theta$, the original symmetry of the
problem is lost, and the geometry is inherently two-dimensional.  This
effect is thus arising from `aliasing' (Kaiser and Peacock 1991,
Szalay \etal 1993, Landy \etal 1996), i.e. projections of the
spherically isotropic power to the lower dimensions of the survey
geometry.  Interestingly enough, even though the power spectrum is
definitely non-local, the redshift space correlation function can
still be computed in spherical coordinates, and it can be expressed in
a closed form, as will be shown below.

In the following Section 2.2, we will first derive a representation of
the redshift distortion problem in terms of spherical tensors. This
development, being a representation independent expansion, is
independent of our choice of coordinate system described above.  In
Section 2.3, we will reintroduce our coordinates and show how the
coefficients of this expansion can be most economically expressed
using these coordinates.

\subsection{Expansion in Spherical Tensors}

We can assume without a loss of generality, that the observer is at
rest with respect to the CMB frame. As shown by HC96, this effect can
be trivially included. The linear expansion of the overdensity at a
redshift space coordinate $\s$ relates to the real-space overdensity
at $\r$, with $\s=\r+u\hr$, as a function of the radial velocity $u$
and its line-of-sight derivative $\partial u/\partial r$, as first
introduced by Kaiser (1987),
\begin{equation}
  \delta^{(s)}(\r) = \delta^{(r)}(\r) -
  \frac{\partial u(\r)}{\partial r} -
  \alpha(r) u(\r),
\label{eq:a1}
\end{equation}
Here $\alpha(r)= (2 + \partial\ln\phi(r)/\partial\ln r)/r$, where
$\phi(r)$ is the selection function which is a slowly varying function
of $r$, the radial distance from the observer.  Due to the fact that
the velocity scale is much smaller than the typical depth of todays
redshift surveys, this term is very small and is generally ignored. We
will include this term for completeness in our results but would like
to make one other point why this term is additionally small.  If the
redshift survey is defined by a boundary on the sky, a given selection
function, and a range of the radial coordinate $(a,b)$, then $\alpha$
will be averaged, weighted by the selection function. Integrating by
parts one can show that this term is small and becomes zero as the
lower limit becomes 0 and the upper limit $\infty$, the case of full
surveys.

The peculiar velocity can be written concisely as a Fourier integral,
with the usual $\beta=\Omega_0^{0.6}/b$, as
\begin{equation}
  u(\r) = i\beta\kint \frac{\hkr}{k} \eikr \tilde{\delta}(\k).
  \label{eq:a3}
\end{equation}
In order to simplify subsequent calculations, we introduce here the
spherical tensor $A$ as
\begin{equation}
  A_l^n(\r) = \kint (ik)^{-n} P_l(\hkr)\eikr \tilde{\delta}(\k).
  \label{eq:a4}
\end{equation}
One can then reexpress the redshift-space overdensity at the point $\r$
in terms of $A$ as
\begin{equation}
  \delta^{(s)}(\r) = \left(1 + \frac{\beta}{3}\right) A_0^0(\r) +
  \frac{2}{3} \beta A_2^0(\r) + \beta \alpha(r) A_1^1(\r).
  \label{eq:a5}
\end{equation}
This use of spherical coordinates makes it easy to express the correlation
of the different $A$ components, at positions $\r_1$ and $\r_2$ as
\begin{equation}
  S_{l_1 l_2}^{n_1 + n_2} =
  \avr{A_{l_1}^{n_1}(\r_1) A_{l_2}^{n_2}(\r_2)}=
  (-1)^{l_2} \kint (ik)^{-(n_1+n_2)}
  P_{l_1}(\hkr_1) P_{l_2}(\hkr_2) \eikr P(k).
 \label{eq:a6}
\end{equation}
where $P_l$ is the Legendre polynomial, $\r = \r_1 -
\r_2$ and $P(k)$ is the power spectrum with the definition
$\avr{\tilde{\delta}(\k)\tilde{\delta}(\k')}
  = (2\pi)^3 \delta^3(\k + \k') P(k)$.
The redshift-space two-point correlation function of finite-angle is
given in terms of $S_{l_1 l_2}^n$, where we will use $\alpha_i=\alpha(r_i)$.
\begin{eqnarray}
  \xi^{(s)}(\r_1,\r_2) &\!\!\!=&\!\!\!
  \left( 1 + \frac{\beta}{3} \right)^2 S_{00}^0 +
  \frac{4}{9} \beta^2 S_{22}^0 +
  \frac{2}{3} \beta \left( 1 + \frac{\beta}{3} \right)
  \left( S_{02}^0 + S_{20}^0 \right)
\nonumber\\
  &+\,& \beta\left( 1 + \frac{\beta}{3} \right)
  \left[ \alpha_1 S_{10}^1 + \alpha_2 S_{01}^1 \right] +
     \frac{2}{3} \beta^2\left[ \alpha_1 S_{12}^1 +
     \alpha_2 S_{21}^1 \right]
     -\,\beta^2 \alpha_1\alpha_2 S_{11}^2
 \label{eq:a8}
\end{eqnarray}
Both the Legendre polynomials and the plane wave $\eikr$ can be expanded
in terms of spherical harmonics as
\begin{eqnarray}
   P_l(\hkr) &\!\!\!=&\!\!\!
   \frac{4\pi}{\qq{l}}\sum_{m = -l}^{l} Y_{lm}^{*}(\hk) Y_{lm}(\hr),
\label{eq:a9}\\
   \eikr &\!\!\!=&\!\!\!
   4 \pi \sum_{l = 0}^\infty \sum_{m = -l}^{l}
   i^l j_l(kr) Y_{lm}^{*}(\hk) Y_{lm}(\hr),
 \label{eq:a10}
\end{eqnarray}
where $j_l$ and $Y_{lm}$ are spherical Bessel functions and
spherical harmonics, respectively. We can express the integral of
three $Y_{lm}$'s over $d\mOmega_\k$ with the Wigner $3j$-symbols.  We
also introduce the bipolar spherical harmonics $X^{LM}_{l_1l_2}
(\hr_1,\hr_2)$ which transform as a spherical harmonic with $L,M$
with respect to rotations (Varshalovich \etal 1988).
\begin{equation}
  X^{LM}_{l_1l_2}(\hr_1,\hr_2) = (-1)^{l_1-l_2-M} \sqrt{\qq{L}}
       \sum_{m_1,m_2}\threej{l_1}{l_2}{L}{m_1}{m_2}{-M}
     Y_{l_1m_1}(\hr_1) Y_{l_2m_2}(\hr_2)
  \label{eq:a11}
\end{equation}
With $Y_{LM}(\hr)$ these can form a rotationally invariant scalar,
dependent on $L, l_1, l_2$, where $\Delta$ indicates the shape of the
triangle formed by the three unit vectors. Any quantity that is a
scalar function of $\Delta$ is independent of what angles or
parameters we may choose to describe the triangle shape. This
invariant scalar $B$ is symmetric in $l_1,l_2$.
\begin{equation}
  B^L_{l_1l_2}(\Delta) =
        {1\over\sqrt{(\qq{l_1})(\qq{l_2})}}
        \threej{l_1}{l_2}{L}{0}{0}{0}
        \sum_M X^{LM*}_{l_1l_2}(\hr_1,\hr_2)Y_{LM}(\hr).
\end{equation}
With this notation, equation (\ref{eq:a6}) reduces to
\begin{equation}
  S_{l_1 l_2}^n =  (4\pi)^{3/2}(-1)^{l_1}
     \sum_L  i^{L-n}
     B^L_{l_1l_2}(\Delta)\, \xi_L^{(n)}(r)
  \label{eq:a12}
\end{equation}
where
\begin{equation}
  \xi_L^{(n)}(r) = \frac{1}{2\pi^2}
  \int dk\,k^2 k^{-n} j_L(kr) P(k),
\label{eq:a13}
\end{equation}

\subsection{Expansion of the Correlation Function}

The above expression shows the correlation function can be written in
terms of a series, factorized into size $r$ and shape $\Delta$
dependent terms, providing a representation independent expansion.
Equations (\ref{eq:a8}) and (\ref{eq:a12}) give an expression of the
redshift-space correlation function. Re-expanding in terms of
$\xi^{(n)}_L(r)$, the $\Delta$ dependence is contained in the
coefficients $c_{ij}(\Delta)$:
\begin{equation}
  \xi^{(s)}(r, \Delta) =  c_{00} \xi^{(0)}_0 + c_{02} \xi^{(0)}_2 +
  c_{04} \xi^{(0)}_4 + c_{11} \xi^{(1)}_1 +\, c_{13} \xi^{(1)}_3 +
  c_{20} \xi^{(2)}_0 + c_{22} \xi^{(2)}_2,
 \label{eq:a16}
\end{equation}
It is at this point that the use of any particular coordinate system
becomes paramount.  Since the coefficients $c_{ij}(\Delta)$ are
functions of the shape of the triangle under consideration, a
judicious use of coordinates can simplify the problem considerably, as
will be shown below.  The terms proportional to $\sin^2(\theta)$ have
been saparated, since these disappear in the plane-parallel
limit. After a somewhat tedious reduction procedure, most of it
carried out in Mathematica, we obtain the following results:
\begin{eqnarray}
  c_{00} &\!\!\!=&\!\!\!
  1 + \frac23 \beta + \frac{1}{5} \beta^2 -
  \frac{4}{15} \beta^2 \cos^2\theta \sin^2 \theta ,
\label{eq:a47}\\
  c_{02} &\!\!\!=&\!\!\!
  - \left( \frac43 \beta + \frac47 \beta^2 \right)
  \cos 2\theta P_2(\cos\gamma)
  - \frac23 \left(\beta - \frac17 \beta^2 +
  \frac47 \beta^2 \sin^2\theta \right) \sin^2\theta,
\label{eq:a48}\\
  c_{04} &\!\!\!=&\!\!\!
  \frac{8}{35} \beta^2 P_4(\cos\gamma) -
  \frac{4}{21} \beta^2 \sin^2 \theta P_2(\cos\gamma) -
  \frac{1}{5} \beta^2 \left(\frac{4}{21} - \frac{3}{7} \sin^2 \theta \right)
  \sin^2 \theta
\label{eq:a49}
\end{eqnarray}
and
\begin{eqnarray}
  c_{11}& =\,\,\,&
  (\alpha_1 - \alpha_2)
  \left(
    \beta + \frac35 \beta^2 - \frac45 \beta^2 \sin^2\theta
  \right)
  \cos\theta P_1(\cos\gamma)
\nonumber\\
  &\,\,\, + &
  (\alpha_1 + \alpha_2)
  \left(
    \beta + \frac35 \beta^2 - \frac45 \beta^2 \cos^2\theta
  \right)
  \sin\theta P_1(\sin\gamma)
\label{eq:a43}\\
  c_{13} &=\,\,\,&
  \frac15 (\alpha_1 - \alpha_2) \beta^2
  \cos\theta
  [\sin^2 \theta P_1(\cos\gamma) - 2 P_3(\cos\gamma)]
\nonumber\\
  &\,\,\, +&
  \frac15 (\alpha_1 + \alpha_2) \beta^2
  \sin\theta
  [\cos^2 \theta P_1(\sin\gamma) - 2 P_3(\sin\gamma)]
\label{eq:a44}\\
  c_{20} &=\,\,\,&
  \frac19 \alpha_1 \alpha_2 \beta^2 (4 \cos\theta - 1),
\label{eq:a45}\\
  c_{22} &=\,\,\,&
  \frac13 \alpha_1 \alpha_2 \beta^2 \sin^2\theta -
  \frac23 \alpha_1 \alpha_2 \beta^2 P_2(\cos\gamma).
\label{eq:a46}
\end{eqnarray}
The first set of coefficients, which do not contain $\alpha$, capture
most of the relevant physics. For the sake of
completeness, the second set gives the $\alpha$-dependent terms $c_{11}$, $c_{13}$,
$c_{20}$ and $c_{22}$ which must be included for calculations in which $\alpha_1$ and
$\alpha_2$ cannot be neglected.

     From these expressions, it is evident that a good choice of coordinate system matters
a great deal. Using the system discussed above (angles $\theta$ and
$\gamma$, separation $r$), leads to remarkably simple expressions for
the coefficients $c$. Other systems we explored generally
contained a lot of associated Legendre polynomials.
It is easily seen in the above expression that taking the limit $\theta \rightarrow 0$
reproduces Kaiser-Hamilton's plane-parallel result.

\section{Discussion}

We have presented a derivation of the redshift space correlation function
between galaxies at two infinitesimal volume elements separated by an arbitrary
angle, and have obtained very simple closed form analytic expressions for the correlations.
These equations show that the effects of a finite angle are quite important. These are
illustrated graphically in Figure 2 which indicates how the redshift distortions change as the
angle between the two lines of sight is increased from 0
to 120 degrees.

Since most of the next generation redshift surveys are wide angle with the majority of pairs
separated at angles greater than 10 degrees, the use of these
expressions will dramatically increase the amount of information which can be extracted
from the surveys resulting in more robust measurements and smaller
shot noise contamination.  Also, if the survey is not contiguous but
consists of several slices like the Las Campanas Survey (Landy etal
1996), or the 2dF survey which will be formed from hundreds of
pencilbeams, this approach can use all the data together to estimate the
infall distortions.

     Another promising approach to study the large scale behavior of the
power spectrum is with building up a Karhunen-Loeve basis (Vogeley and
Szalay 1996). The first step in that method is to subdivide the survey
into small cells (in redshift space) and then construct their correlation
matrix. Most pairs of cells will have a large relative angle. Thus,
with the expressions presented here, computing the correlation matrix
is a trivial exercise. Once the appropriate basis has been created,
one can use the Fisher matrix to select the set of eigenmodes most
sensitive to $\beta$, yielding an optimal estimation of the value of
$\beta$.

     In conclusion, we believe that our results allow a much more elegant treatment
of the redshift distortions in a general geometry than before with the added benefit that all
galaxy pairs may be utilized to construct a signal.

\acknowledgements

 AS would like to acknowledge the hospitality of the University of Tokyo,
where this work has been completed. Useful discussions with Y. Suto are
also acknowledged.

\begin{figure}
\plotone{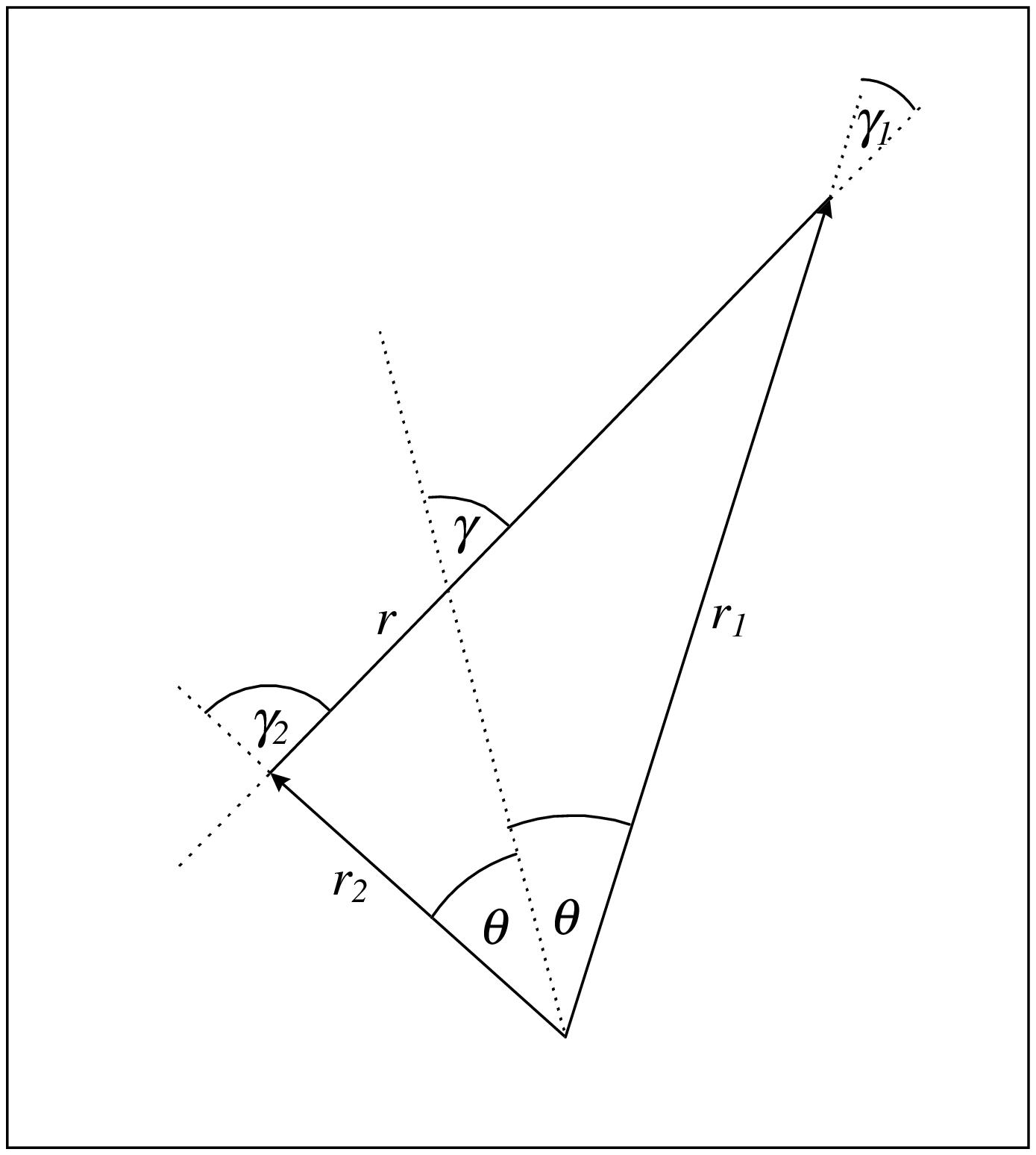}
\figcaption{A graphic representation of our coordinate system.  The angle
between the two lines of sight is $2\theta$. The main line-of-sight is
taken to be the axis bisecting the angle $2\theta$, since this is the
symmetry axis for the two directions $\hr_1$ and $\hr_2$.  The angle
$\gamma$ is defined by the main line of sight and the vector $\r$.
}
\end{figure}

\begin{figure}
\plotone{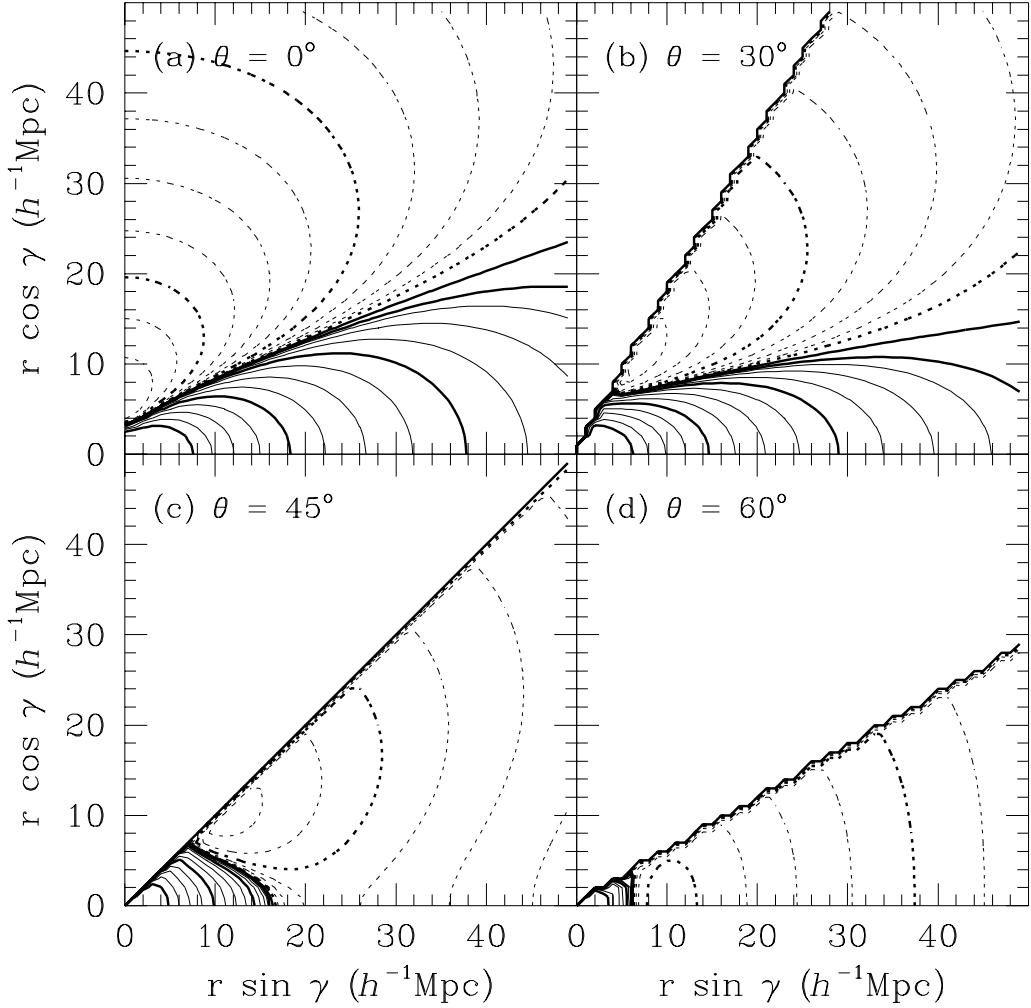}
\figcaption{
Contour plots of the redshift-space correlation functions in polar
coordinate system $r, \gamma$ for fixed finite angles
$\theta$. A standard CDM model with COBE normalization and $H_0 = 70$ km
s$^{-1}$ Mpc$^{-1}$ is assumed. Plots (a), (b), (c) and (d)
correspond to $\theta = 0, 30, 45$ and $60$ degrees. Solid and
dashed lines indicate positive and negative
values, respectively. Contour spacings are $\Delta \log_{10}|\xi| =
0.25$. Solid thick lines represent the value 100, 10, 1 and
0.1. Dashed thick lines represent the value -1 and -0.1. Blank areas
correspond to unphysical regions $\gamma > \theta$.
}
\end{figure}

\end{document}